\documentclass[12pt]{article}
\usepackage{amssymb,epsf,graphicx}

\def\bra#1{\langle #1 |}
\def\ket#1{|#1 \rangle}
\def\aver#1{\langle\, #1 \,\rangle}

\def\l{\left}
\def\r{\right}

\let\eps = \varepsilon
\def \be {\begin{equation}}
\def \ee {\end{equation}}
\def \bea {\begin{eqnarray}}
\def \eea {\end{eqnarray}}
\def \bdm {\begin{displaymath}}
\def \edm {\end{displaymath}}

\def \nn {{\mathbb N}}
\def \zz {{\mathbb Z}}
\def \cc {{\mathbb C}}
\def \rr {{\mathbb R}}

\def \QQ {{\cal Q}}
\def \nnn {{\cal N}}

\def \nc {noncommutative }

\def \sf  {string field }
\def \sft {string field theory }

\begin{document}

\begin{flushright}
MIT-LNS-02-297\\
MIT-CTP-3249\\
{\tt hep-th/0202139}
\end{flushright}

\vskip 2cm

\begin{center}
{\LARGE Anomalous reparametrizations and butterfly states
in string field theory \\}
\vskip 1cm
{\Large Martin Schnabl}\footnote{E-mail: schnabl@lns.mit.edu} \\
\vskip 0.5cm
{\it
Center for Theoretical Physics, Massachusetts Institute of Technology,\\
Cambridge, MA 02139, USA}
\end{center}

\begin{abstract}
The reparametrization symmetries of Witten's vertex in ordinary or
vacuum string field theories can be used to extract useful information
about classical solutions of the equations of motion corresponding to
D-branes. It follows, that the vacuum string field theory in general has to be
regularized. For the regularization recently considered by Gaiotto
et al., we show that the identities we derive, are so constraining, that
among all surface states they uniquely select the simplest butterfly projector
discovered numerically by these authors. The reparametrization
symmetries are also used to give a simple proof that the butterfly
states and their generalizations are indeed projectors.
\end{abstract}

\baselineskip=18pt

\section{Introduction}

The open \sft \cite{Witten:NCGSFT} has so far resisted all the
attempts to find analytically its solutions corresponding to the true
closed string vacuum or lower dimensional D-branes. On the other hand the
solutions are known numerically \cite{SZ,MT} and provide a good
evidence for the Sen's tachyon conjectures \cite{Sen}.
The vacuum \sft (VSFT) \cite{RSZ1,RSZ5},
designed to describe the physics around the true tachyon vacuum has
been conjectured to have purely ghost kinetic term, which then allows
for some exact calculations. Canonical choice of a ghost kinetic term
has been recently proposed by Gaiotto, Rastelli, Sen and Zwiebach
\cite{GRSZ} based on some earlier results of Hata and Kawano \cite{HK}.
It has been further discussed by Okuyama \cite{Okuyama1}.

Generic feature of pure ghost kinetic terms in VSFT is that they allow
for simple construction of solutions in terms of projectors in the
matter part of the \sf algebra. It has been proved that such
solutions indeed have the right ratios of tensions to be interpreted
as D-branes \cite{RSZ4,Okuyama2}. To determine however the absolute
normalization turned out to be quite problematic. As we will show, on
general grounds one should expect some sort of divergences in the
normalization of the action. To deal with the problem, the authors of
\cite{GRSZ} added a small term to the kinetic operator proportional to
$c_0 L_0$. With this regularization they observed that the
numerical solution in the limit of removed regulator approaches a new
type of a projector, called butterfly projector.
The poetic name comes from the shape of the surface which represents
these kind of states in the conformal field theory. Some properties of
these states were also reported in \cite{SenJHS}.
For other very recent development in the VSFT see 
\cite{BMS,GRSZ2,Arefeva,Hata2,Kluson,DLMZ}.

In this paper we would like to get more insight into the problem
of classical solutions and butterfly states in VSFT by an
extension of the analysis of our earlier work \cite{Schnabl2}. The
basic idea is to study implications of anomalous symmetries of the
Witten's vertex for classical solutions of the action. These
techniques turn out to be remarkably useful in this new context.
They show that in the vacuum \sft one inevitably encounters some
sort of singularities. For the regularized version of the theory
we obtain infinite set of nontrivial identities which, as we show,
are fulfilled for the simplest butterfly state. The identities are
so constraining, that among all well defined surface states they
uniquely pick up one particular state, which even happens to be a
projector. It would be rather interesting if one could derive more
identities which would then determine the projector uniquely in
the the whole Hilbert space, not necessarily in the subspace of
surface states.

It turns out to be possible to express any well defined surface state
as a reparametrization of the vacuum in the form
\be\label{UfeK}
\bra{0} U_f = \bra{0} e^K,
\ee
which greatly simplifies calculations of some star products. Here
$K$ stands for some  linear combination of the generators
$K_{n} = L_{n} - (-1)^n L_{-n}$, which are known to act as exterior
derivatives in the \sf algebra. This fact has been shown in a recent
paper \cite{Schnabl3} for wedge states. It allowed to determine the
star products of wedge states purely algebraically, without thinking
much about gluing surfaces and various coordinate systems.

Particularly interesting class of states is the one of the form
\be
\lim_{\alpha \to \pm \infty} \bra{0} e^{\alpha K}.
\ee
If the limit exists, then it gives states which are invariant under
the reparametrization $K$, i.e. they are annihilated by $K$. These
states are natural candidates for being projectors. For example the
simplest butterfly surface state 
\be\label{butt2}
\bra{0} e^{-\frac{1}{2}L_{2}}  =   \lim_{\alpha \to - \infty} \bra{0} e^{\alpha K_2}
\ee
is a projector. On the other hand the opposite limit
\be\label{butt2f}
\bra{0} e^{+\frac{1}{2}L_{2}}  =   \lim_{\alpha \to + \infty} \bra{0} e^{\alpha K_2}.
\ee
is not a projector, we call it a `false butterfly' state.
These kind of states (\ref{butt2},\ref{butt2f}) can be naturally 
generalized by replacing $K_2$ with $K_{2k}$. 

The paper is organized as follows: In section 2 we study the
consequences of the anomalous symmetries of the Witten's three
vertex for the classical solutions of the ordinary, vacuum and
regulated vacuum \sf theories. We derive an infinite set of
identities, which as we show are obeyed by the simplest butterfly
state. We then show in section 3, that the butterfly states and
their generalization are indeed projectors. We manage also to
calculate the star product $e^{s L_{-2}} \ket{0}
* e^{s L_{-2}} \ket{0}$ for arbitrary $s \in
[-\frac{1}{2},\frac{1}{2}]$. In section 4 we prove the uniqueness
of the solution of the set of identities among the surface states.
As a corollary we obtain a formula (\ref{UfeK}). Appendices
contain some useful formulas for the exponentials of the Virasoro
operators, conservation laws for butterfly states and some
numerical results for illustrative purposes.

Related  work on butterfly states with some small overlaps with
our work has been done independently by Gaiotto, Rastelli, Sen and
Zwiebach and should appear simultaneously in \cite{GRSZ3}.

\section{Constraints from anomalous symmetries}
\label{Constraints}

To start with, we shall recall first some identities for the tachyon condensate
found in our earlier work \cite{Schnabl2}.

\subsection{Constraints in the Witten's \sft}

The action of Witten's \sft \cite{Witten:NCGSFT} takes a nice form of \nc
Chern-Simons action
\begin{equation}\label{action}
S[\Phi]=-\frac{1}{\alpha' g_o^2} \left( \frac 12 \aver{\Phi, Q \Phi} +
\frac 13 \aver{\Phi,\Phi*\Phi} \right),
\end{equation}
where $Q$ is the usual string BRST operator. The star multiplication is given by a
three vertex $\bra{V}$  \cite{GJ,Samuel} which satisfies various
identities \cite{GJ,Romans,RZ}. For our purposes the most important are
\bea\label{ansyms}
\bra{V} \sum_{i=1}^3 ( L_{-n}^{X(i)} - L_{n}^{X(i)} ) &=& 3 k_n^x \bra{V},
\nonumber\\
\bra{V} \sum_{i=1}^3 ( L_{-n}^{gh(i)} - L_{n}^{gh(i)} ) &=& -3 k_n^x \bra{V},
\nonumber\\
\bra{V} \sum_{i=1}^3 ( J_{-n}^{(i)} + J_{n}^{(i)} ) &=& 3 ( h_n^{gh} + 3\delta_{n,0}) \bra{V},
\eea
valid for $n$ even and $L_n^X$ and $L_n^{gh}$ denote matter or
ghost Virasoro operators and $J_n$ are ghost current generators.
The constants $k_n^x$ and $h_n^{gh}$ for $n$ even are given
explicitly by
\bea
k_n^x &=& \frac{13 \cdot 5}{27} \cdot \frac{n}{2} (-1)^{\frac n2},
\nonumber\\
h_n^{gh} &=& - (-1)^{\frac n2}.
\eea

Taking the variation of the action (\ref{action}) under the infinitesimal
variations of the \sf $\Phi$
\bea\label{variations}
\delta \Phi &=& ( L_{-n}^X - L_{n}^X -  k_n^x ) \Phi,
\nonumber\\
\delta \Phi &=& ( J_{-n} + J_{n} -  h_n^{gh} - 3\delta_{n,0} ) \Phi
\eea
we see that they leave the cubic term in the action invariant.
On the other hand we know that the total action should also be invariant,
as long as $\Phi$ satisfies equations of motion.
Combining these two facts we get from the kinetic term
\bea\label{anids}
\bra{\Phi} [Q, L_n^X] \ket{\Phi} &=&   -k_n^x \bra{\Phi} Q \ket{\Phi},
\nonumber\\
\bra{\Phi} [Q, J_n] \ket{\Phi} &=&   h_n^{gh} \bra{\Phi} Q \ket{\Phi}.
\eea
These identities in the Siegel gauge simplify to
\bea\label{anidsSiegel}
\bra{\Phi} c_0 L_n^X \ket{\Phi} &=&  \frac 1n k_n^x \bra{\Phi} c_0 L_0^{tot} \ket{\Phi},
\nonumber\\
\bra{\Phi} c_0 (n J_n + L_n^{tot}) \ket{\Phi} &=&   -h_n^{gh} \bra{\Phi} c_0 L_0^{tot} \ket{\Phi},
\eea
where $L_n^{tot}=L_n^X+L_n^{gh}$ is the total Virasoro generator. They can be further rewritten as
\bea\label{anidsSiegel2}
\bra{\Phi} c_0 L_n^X \ket{\Phi} &=&  + \frac{65}{54} (-1)^{\frac{n}{2}} K,
\nonumber\\
\bra{\Phi} c_0 {L'}_{n}^{gh} \ket{\Phi} &=& -\frac{11}{54} (-1)^{\frac{n}{2}} K,
\eea
where  $K=\bra{\Phi} c_0 L_0^{tot} \ket{\Phi}$ and 
${L'}_{n}^{gh}=L_n^{gh}+n J_n + \delta_{n,0}$ are Virasoro
operators of the twisted ghost conformal
field theory \cite{GRSZ} with central charge $c=-2$. Note that the operators
${L'}_{n}^{gh}$ are singlets under the $SU(1,1)$ symmetry in the Siegel gauge, discussed in
\cite{Zwiebach}.

\subsection{Constraints in the vacuum \sft}
\label{ConstrVSFT}

The basic idea of the vacuum \sft (VSFT) is that when the Witten's \sft is expanded
around the nonperturbative vacuum, its action can be again written in the form
\begin{equation}\label{VSFTaction}
S[\Psi]=-\kappa_0 \left( \frac 12 \aver{\Psi, \QQ \Psi} +
\frac 13 \aver{\Psi,\Psi*\Psi} \right),
\end{equation}
where $\Psi$ is the fluctuation around the new vacuum after suitable reparametrization
and $\QQ$ is some new BRST-like operator which is nilpotent, acts as a derivation of the \sf
algebra and is conjectured to be purely ghost. Recently it has been
argued \cite{GRSZ,Okuyama1},
that there are certain singular reparametrizations which lead
naturally to one particular kinetic
operator
\be\label{Q-GRSZ}
\QQ= c_0 + \sum_{k=1}^\infty (-1)^k (c_{2k} + c_{-2k}).
\ee

The major advantage of VSFT with the purely ghost kinetic operator
$\QQ$ is that it allows for constructing D-branes in terms of
projectors of the algebra. They have been shown exactly, using the
methods of boundary conformal field theory and Fock space oscillators, 
to have the right ratios of tensions \cite{RSZ4, Okuyama2}. 
What is still missing, is to show that there
is a solution corresponding to the original unstable D25-brane
with the right tension. We will see that the analysis of the
constraints analogous to those discussed in the previous subsection
can shed some light on the nature of this D25-brane solution.

Let $\Psi_0$ be a stationary point of the action (\ref{VSFTaction}), i.e. a solution
of the equations of motion $\QQ \Psi_0+ \Psi_0*\Psi_0=0$. Then the identities read
\bea\label{anidsVSFT}
\bra{\Psi_0} [\QQ, L_n^X] \ket{\Psi_0} &=&   -k_n^x \bra{\Psi_0}
\QQ \ket{\Psi_0},
\nonumber\\
\bra{\Psi_0} [\QQ, L_n^{gh}] \ket{\Psi_0} &=&   k_n^x \bra{\Psi_0}
\QQ \ket{\Psi_0},
\nonumber\\
\bra{\Psi_0} [\QQ, J_n] \ket{\Psi_0} &=&   h_n^{gh} \bra{\Psi_0}
\QQ \ket{\Psi_0}.
\eea
For a generic $\QQ$ which is purely ghost, the first equation
implies that
\be
\bra{\Psi_0} \QQ \ket{\Psi_0} = 0.
\ee
It seems therefore to show that the tension of the D25-brane
\be
\frac{\kappa_0}{6} \bra{\Psi_0} \QQ \ket{\Psi_0}
\ee
and of any other soliton is zero. The same conclusion follows for the particular choice
(\ref{Q-GRSZ}) also from a combination of the second and third identity of
(\ref{anidsVSFT}) in the Siegel gauge.

So, does this mean that the VSFT cannot describe finite tension
D25-branes? No, the argument merely shows that for the soliton
solutions things become singular and one must be very careful. In
the scenario advocated in \cite{GRSZ} for the emergence of purely
ghost kinetic operator in VSFT, the constant $\kappa_0$ in front
of the action becomes naturally singular. When the action is
carefully regularized we may hope to get the right finite answer.

One might be curious whether the above argument definitely rules out the possibility that one
day we find completely regular VSFT action with purely ghost kinetic term and with finite action
solutions. As a possible example of what could happen, the following toy version of VSFT has been
proposed by \cite{GRSZ,RSZprivate}. Suppose we start with a string
field of ghost number zero and consider an operator
$Y(z)=\frac 12 c\partial c \partial^2 c(z)$, which is a dimension zero primary of ghost number three.
Take an action of the form
\be\label{ghn0sft}
S=\beta \l[ \frac 12 \bra{\Phi} (Y(i)+Y(-i)) \ket{\Phi} +
\frac 13 \bra{\Phi} (Y(i)+Y(-i)) \ket{\Phi * \Phi}\r].
\ee
From the argument in \cite{GRSZ} it follows that for projectors which are surface states in the
total Virasoro algebra (the sliver for example), the action is finite and equal to $\frac 13 \beta$.
So where does the argument breaks down? It is quite simple to find few places in the case of the
butterfly projector
\be
\ket{B} = e^{-\frac12 L_{-2}} \ket{0},
\ee
discovered in \cite{GRSZ} and discussed in more detail in section
\ref{Butterflies} and in \cite{SenJHS,GRSZ3}.
Easy check shows, that for example the conservation laws (\ref{ansyms}) break down.
Indeed, for $n=2$
\bea
&& \bra{V} \sum_{j=1}^3 ( L_{-2}^{X(j)} - L_{2}^{X(j)} ) \, (Y(i)+Y(-i))
\ket{B^{(1)}} \otimes \ket{B^{(2)}} \otimes \ket{B^{(3)}} = \frac{3c}{4}
\nonumber\\
&& \neq
3 k_2^x \bra{V} (Y(i)+Y(-i))\ket{B^{(1)}} \otimes \ket{B^{(2)}} \otimes \ket{B^{(3)}} =
- \frac{5c}{18},
\eea
where $c=26$ is the central charge in the matter sector.
In  section \ref{Butterflies} we will see that some things are actually
ill defined, for instance the inner product $\aver{B|(Y(i)+Y(-i))
L_{-2}^X B}$ and even the star product
$\ket{B} * L_{-2}^X \ket{B}$ itself.

To conclude, our argument rules out possibility of finite
action solutions which are sufficiently regular, it allows however for
less well behaved solutions as are the projectors.

\subsection{Constraints in the regularized VSFT}
\label{ConstrRegVSFT}

In \cite{GRSZ} the following regularized action of the VSFT in the Siegel gauge
has been proposed
\be\label{regVSFTaction}
S_a[\Psi]=-\kappa_0(a) \left( \frac 12 \aver{\Psi, c_0 (1 + \frac 1a L_0^{tot}) \Psi} +
\frac 13 \aver{\Psi,\Psi*\Psi} \right).
\ee
The original gauge fixed VSFT is recovered in the limit $a \to \infty$. It is perhaps curious
that the other limit $a \to 0$ corresponds to the ordinary Witten's
\sf theory. One might therefore
hope that both actions share some features, and that is indeed true.
Let us write the constraints on the solutions to the equation of motion. Since we are in the
Siegel gauge, we may write only identities based on variations $\delta\Psi$, which preserve
the gauge
\bea\label{anidsregVSFT}
\bra{\Psi_0} c_0 L_n^X \ket{\Psi_0} &=&  \frac{65}{54} (-1)^{\frac{n}{2}} a K,
\nonumber\\
\bra{\Psi_0} c_0 (n J_n + L_n^{gh}) \ket{\Psi_0} &=&
-\frac{11}{54} (-1)^{\frac{n}{2}} a K,
\eea
where
\be K = \bra{\Psi_0} c_0(1 +\frac 1a  L_0) \ket{\Psi_0}.
\ee
It was observed in \cite{GRSZ} that in the limit $a \to \infty$
the numerical solution of the equations of motion representing the
D25-brane seems to approach the following state
\be\label{GRSZsol}
\ket{\Psi_0} = \nnn\, e^{-\frac12 (L_{-2}^X + {L'}_{-2}^{gh})} c_1
\ket{0}.
\ee
We will now show that this solution is compatible with the
infinite set of identities (\ref{anidsregVSFT}) even for finite $a$. 
An immediate question is then, whether the solution (\ref{GRSZsol}) is a unique
solution of this set of identities. We will prove in section 
\ref{Uniqueness}, that it is indeed unique among the surface
states in the deformed CFT. It is most likely not unique among
non-surface states, since for example in the limit $a\to 0$
corresponding to ordinary \sf theory, we have another solution. 
To deal with non-surface states, one would almost certainly 
need more identities, which would differ from (\ref{anidsregVSFT}) by
taking expectation values of products of more operators.

To check the identities we have to first regularize the solution (\ref{GRSZsol})
\be
\ket{\Psi_0^{s, \tilde s}} = \nnn\,  e^{ s L_{-2}^X + \tilde s
{L'}_{-2}^{gh}} c_1 \ket{0}
\ee
keeping $|s|$ and $|\tilde s| < \frac 12$,
and let $s$ and $\tilde s$ approach $-\frac 12$. 
Notice, that the regularized solution is not necessarily a solution of
the regularized action. We are taking fixed finite $a$ and want to
show that the identities are obeyed in the limit $s,\tilde s \to -\frac 12$.

Let us focus first on the identity
\be
\frac{\bra{\Psi_0} c_0 L_2^X \ket{\Psi_0}}{\bra{\Psi_0} c_0
{L'}_{2}^{gh} \ket{\Psi_0}} = -\frac{65}{11},
\ee
which follows from (\ref{anidsregVSFT}). 
By a straightforward calculation, whose details are given in Appendices
\ref{AppTransf} and \ref{RegGRSZ}, we see that we have
to require that in our limit 
\be\label{fixedratio}
\frac{1-4{\tilde s}^2}{1-4s^2} = \frac{5}{11}
\ee
is kept fixed. Remarkably, making this single adjustment suffices to
show that all other identities (\ref{anidsregVSFT}) are satisfied.
Using the formulas collected in Appendix \ref{RegGRSZ} we get
\be\label{aK}
aK = \bra{\Psi_0}(a c_0 + c_0 L_0^{tot}) \ket{\Psi_0} \approx
\frac{27}{5} \l(\frac{5}{11}\r)^{\frac{1}{4}} \nnn^2 \frac{1}{(1-4s^2)^{3}}.
\ee
Here it is important, that we take first the limit $s,\tilde s \to -\frac 12$
for fixed $a$, and only at the end when we get $a$ independent
ratios we may send $a \to \infty$, if we wish so. We have also
\bea\label{buttid}
\bra{\Psi_0} c_0 L_{2k}^X \ket{\Psi_0} &\approx& (-1)^k \frac{26}{4}
\l(\frac{5}{11}\r)^{\frac{1}{4}} \nnn^2 \frac{1}{(1-4s^2)^{3}},
\nonumber\\
\bra{\Psi_0} c_0 {L'}_{2k}^{gh} \ket{\Psi_0} &\approx & -(-1)^k \frac{11}{10}
\l(\frac{5}{11}\r)^{\frac{1}{4}} \nnn^2 \frac{1}{(1-4s^2)^{3}}.
\eea
The identities (\ref{anidsregVSFT}) now immediately follow from
(\ref{aK}) and (\ref{buttid}). 

Let us make one more remark about the logic here. We have shown that
with a finite regularization $a$ of the VSFT action, the butterfly
state exactly solves the infinite set of identities. In order to prove that,
we had to regularize the state as we did. We have not shown, that it
exactly obeys the equations of motion of the regularized theory. We
will see in section \ref{DetailedButt} however, that the regularized butterfly
approximately solves the equations of motion of the regularized VSFT,
provided we identify $s=\tilde s$ and let it depend on $a$ in a definite way.


\section{Butterfly states}
\label{Butterflies}

Butterfly states were introduced first in \cite{GRSZ,SenJHS} as
surface states associated to the conformal mappings\footnote{Our
parameter $r$ is related to the parameter $\alpha$ of Gaiotto et
al. \cite{GRSZ} by $r=\frac{2}{\alpha}$.}
\bea\label{defbuttmap}
f_r(z) &=&  \sin\l(\frac{2}{r} \arctan z\r)
\nonumber\\
&=& \frac{z}{\sqrt{1+z^2}} \circ \tan\l(\frac{2}{r} \arctan z\r).
\eea
Various descriptions of the surface states have been developed
recently in \cite{RSZ4}. Let us review some definitions and
properties we will need later. For a conformal field theory (CFT)
defined on upper half plane, take a map $f(z)$ which maps real
axis to real axis and the upper unit half disk one to one in the
upper half plane. The surface state associated to the map $f$ is
defined as a linear functional on the Hilbert space, which is just
a bra vector, satisfying
\be\label{defsurf}
\aver{f|\phi} = \aver{0|\, U_f\, |\phi} =\aver{f \circ \phi(0)},
\qquad \forall \phi.
\ee
Here $\phi(0)$ is the operator corresponding to the state
$\ket{\phi}$ through the operator--state correspondence, and $f
\circ \phi(0)$ denotes its conformal transformation. The operators
\be
U_f = e^{\sum_{n\ge 0} v_n L_n}
\ee
satisfying 
\be
f \circ \phi(z) = U_f \phi(z) U_f^{-1}, \qquad \forall \phi
\ee 
form a representation of the conformal group on the Hilbert space
of the worldsheet CFT. The coefficients $v_n$ in front of the Virasoro
operators $L_n$ can be determined by solving certain recursive
equations. The vector field $v(z) = \sum v_n z^{n+1}$ is also
determined up to a constant by the equation $v(z)\partial f =
v(f(z))$. For recent review and discussion of related issues the
reader is referred to \cite{Schnabl3}.

\begin{figure}[ht]
\begin{center}
\input{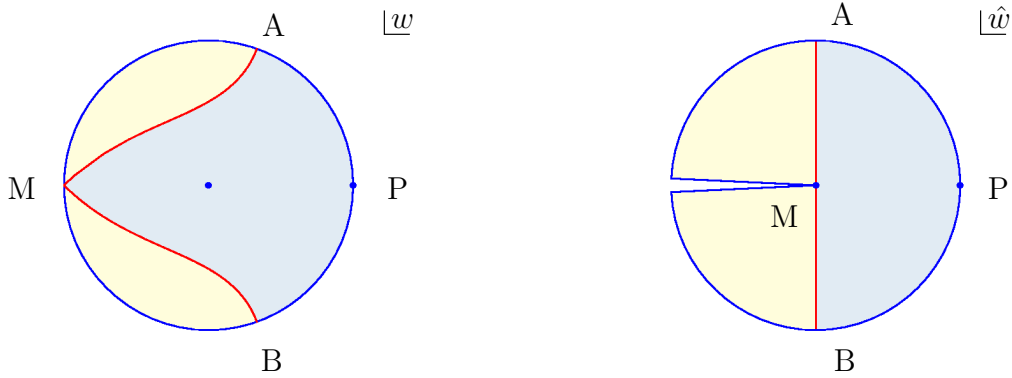}
\caption{\small Representation of the butterfly state $\bra{0}
e^{-\frac12 L_{2}}$ in two coordinate systems. The local coordinate
patch around the puncture $P$ is indicated by the darker shade.
In the $\hat w$ coordinate system the local patch is
represented always as the right half disk. The rest of the surface in
this case is a half disk cut into two parts along the line $[-1,0]$.
}
\label{FigButt2}
\end{center}
\end{figure}

Using conformal map
\be
h(z) = \frac{1+i z}{1-i z},
\ee
which maps the upper half plane with a coordinate $z$ one to one onto
the unit disk with a coordinate $w=h(z)$ we can evaluate the
correlator (\ref{defsurf}) as
\be
\aver{f \circ \phi(0)}_{\rm UHP} = \aver{h\circ f \circ
\phi(0)}_{\rm disk}.
\ee
The image of the unit upper half disk under the map $h \circ f$ is
called local coordinate patch. For the simplest butterfly surface
state corresponding to the map (\ref{defbuttmap}) with $r=2$ the
shape of the local patch is indicated in Fig. \ref{FigButt2}.
Recall, that in this coordinate system the local coordinate patch
for the sliver state looks just as a sliver.

There is another coordinate system with a coordinate 
$\hat w= h \circ f^{-1} \circ h^{-1}(w)$, 
in which the local patch is mapped to the right half
disk. In this coordinate system the star product of surface states is
represented simply by gluing the surfaces with the local patch
detached. For the butterfly state, the part of the surface without the 
local patch looks as a butterfly, hence the name. For the sliver the
surface in this coordinates looks as a helix of an
infinite excess angle.

For general $r$, the butterfly states can be expressed explicitly
as elements of the Verma module (representation space of the Virasoro
algebra)
\be\label{Buttr2}
\ket{B,r} = U_r^\dagger e^{-\frac{1}{2} L_{-2}}\ket{0},
\ee
where $U_r^\dagger$ are well known from the studies of wedge states
\cite{RZ,Schnabl3}. They are given by
\be
U_r^\dagger = e^{2\log\frac{r}{2} A^\dagger},
\ee
where
\be\label{defA}
A= -\frac 12 L_0 + \sum_{k=1}^\infty \frac{(-1)^k}{(2k-1)(2k+1)}
L_{2k}
\ee
and the dagger denotes the BPZ conjugation $L_n^\dagger = (-1)^n L_{-n}$.
Similar vector field has been also considered recently in \cite{RSZ7}.

\begin{figure}[t]
\begin{center}
\input{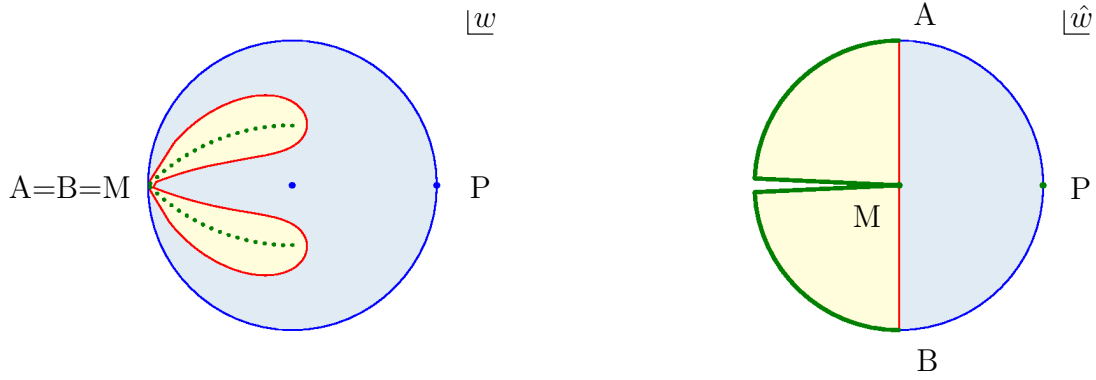}
\caption{\small Representation of the butterfly state $\bra{0}
e^{\frac14 L_{4}}$ in two coordinate systems.
In the $\hat w$ coordinate system the local patch is
represented by the right half disk, the rest of the surface on the
right is cut along the line $[-1,0]$. Moreover the upper left
$90^\circ$ arc is identified with the upper part of the cut
and the lower left arc with the lower part of the cut.
}
\label{FigButt4}
\end{center}
\end{figure}

From the proof we are going to present in the next subsection it follows,
that the family of butterfly projectors can be generalized also to include
surface states
\be
\ket{B^{(2k)}} = e^{\frac{(-1)^k}{2k} L_{-2k}}\ket{0}
\ee
associated to conformal maps
\be
f(z)=\frac{z}{\sqrt[2k]{1-(-1)^k z^{2k}}}.
\ee
considered already in  \cite{LPP2}. An example of the corresponding
surface for $k=2$ is given at Fig. \ref{FigButt4}.

These are direct generalizations of the state (\ref{Buttr2})
for $r=2$. Once we establish that the states with  $r=2$ are
projectors, then the case $r \ne 2$ follows by considering finite
reparametrizations and will be discussed separately in subsection
\ref{Family}.

\subsection{Proof that butterflies are projectors}
\label{Proof}

Let us assume for simplicity, that our CFT has overall central
charge $c=0$. The complications for nonzero central charge, as in
the matter plus twisted ghost CFT can be dealt with separately.
We start then with the following simple identity
\be
\label{etoK2k}
e^{\alpha K_{2k}} = e^{-\frac{1}{2k}\tanh 2k\alpha\, L_{-2k}}\,
 \l(\cosh 2k\alpha\r)^{- \frac{1}{k} L_0}
 \, e^{\frac{1}{2k} \tanh 2k\alpha\, L_{2k}}
\ee
discussed more in  Appendix \ref{AppGroup}. Using this formula we can
write the butterfly state in a regulated form as
\be
e^{(-)^{k+1}\alpha K_{2k}} \ket{0} = e^{\frac{(-1)^k}{2k}\tanh 2k\alpha\, L_{-2k}} \ket{0},
\ee
where $\alpha \to \infty$. The advantage of this expression for the butterfly is
that $K_{2k}$ are (exterior) derivations of the star algebra and thus
we have in general
\be\label{transfstar}
e^K ( \ket{\phi} * \ket{\chi} ) = e^K  \ket{\phi} *  e^K \ket{\chi},
\ee
for any two string fields $\phi$ and $\chi$.

To prove that the butterfly is a projector we can write
\be
e^{\frac{(-1)^k}{2k} L_{-2k}}\ket{0} * e^{\frac{(-1)^k}{2k}
L_{-2k}}\ket{0} = \lim_{\alpha \to \infty}
e^{(-)^{k+1}\alpha K_{2k}} \ket{0} * e^{(-)^{k+1}\alpha K_{2k}} \ket{0} =
\lim_{\alpha \to \infty} e^{(-)^{k+1}\alpha K_{2k}} U_3^\dagger \ket{0}.
\ee
Now formally holds
\be\label{reparU3}
\lim_{\alpha \to \infty} e^{(-)^{k+1}\alpha K_{2k}} U_3^\dagger \ket{0}=
\lim_{\alpha \to \infty} e^{\frac{(-1)^k}{2k} L_{-2k}}
\l(\cosh 2k\alpha\r)^{- \frac{1}{k} L_0} e^{-\frac{(-1)^k}{2k} L_{2k}}
 U_3^\dagger \ket{0} = e^{\frac{(-1)^k}{2k} L_{-2k}}\ket{0},
\ee
since all the terms in $e^{-\frac{(-1)^k}{2k} L_{2k}} U_3^\dagger
\ket{0}$ with nonzero level get multiplied with vanishing factors.
There is one important potential flaw in this argument however. We could
have repeated all the argument without bothering about the
$(-1)^k$ factor.\footnote{We learned about the subtleties related
to this factor from Barton Zwiebach.} This would imply the
incorrect statement that $e^{\frac{1}{2} L_{-2}} \ket{0}$ is also
a projector. What goes wrong in this case is that $e^{-\frac{1}{2}
L_{2}} U_3^\dagger \ket{0}$ is divergent in the level expansion.
When regulated, the divergences are just cancelled by the
vanishing factors and we end up with something finite.

To complete the proof we have to show that
\be\label{ButtInvWedge}
\lim_{\alpha \to \infty} e^{-\frac{(-1)^{k}}{2k}\tanh 2k\alpha\, L_{2k}}
U_3^\dagger \ket{0} = e^{-\frac{(-1)^{k}}{2k}\, L_{2k}}
U_3^\dagger \ket{0}
\ee
has finite unambiguous limit. For $k=1$ it can be proved by explicit
calculation (see below), for general $k$ we have to use the gluing theorem.

In general for two conformal maps $f$ and $g$ the state $\bra{0} U_f U_g^\dagger$
is again a surface state. It can be calculated as follows: From the
surfaces corresponding to the states $\bra{0} U_f $ and $\bra{0} U_g$
in the $\hat w$ coordinate we detach the local coordinate patches and
glue the remaining surfaces  together along the segment $ABM$.
The part of the surface corresponding to $\bra{0} U_g$ represents now new coordinate
patch and we may map it back to the canonical half-disk. The resulting
surface is then the surface representation of the state $\bra{0} U_f U_g^\dagger$.
This  procedure has been applied to the case of wedge states in
\cite{Schnabl3}, where it was shown to agree with other methods.

\begin{figure}[t]
\begin{center}
\input{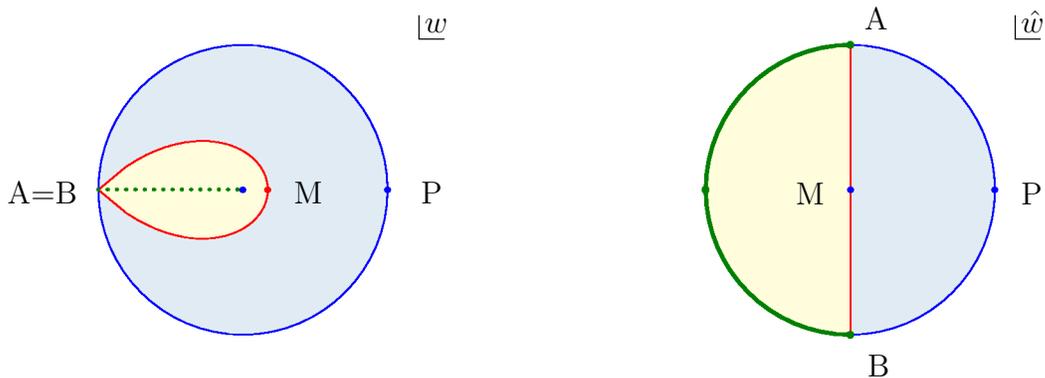}
\caption{\small Representation of the `wrong sign' butterfly
state $\bra{0} e^{\frac12 L_{2}}$.
In the $\hat w$ coordinate system the upper and lower halves of the left
semicircle, indicated by thick line, are identified.
Note that the surface is completely regular
at the midpoint.
}
\label{FigButtInv2}
\end{center}
\end{figure}

\begin{figure}[t]
\begin{center}
\input{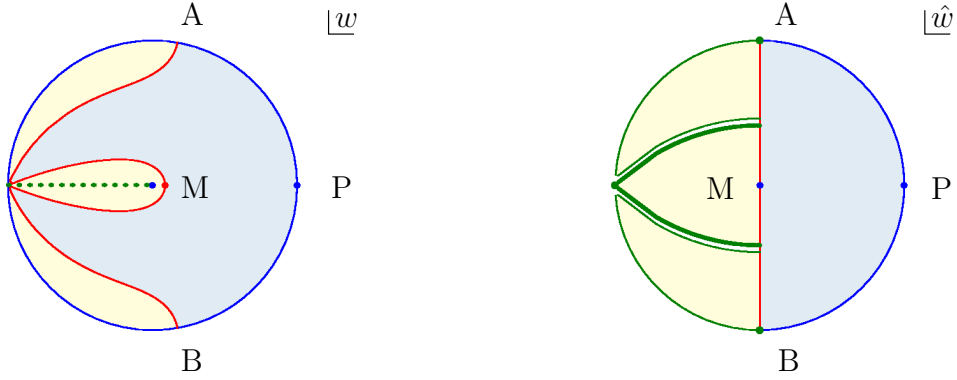}
\caption{\small Representation of the `wrong sign' butterfly
state $\bra{0} e^{-\frac14 L_{4}}$.
In the $\hat w$ coordinate system the surface without the local patch
is split into three independent parts, the middle part has its
boundaries identified. Again the surface is perfectly regular at the midpoint.
}
\label{FigButtInv4}
\end{center}
\end{figure}

In our present case we have to glue a wedge with opening angle $2\pi$
(after detaching the local coordinate patch)  with a surface associated to
the map
\be\label{wrongbuttmap}
f(z)=\frac{z}{\sqrt[2k]{1+(-1)^k \tanh(2k\alpha) \,z^{2k}}}.
\ee
Two examples of these surfaces are in Fig. \ref{FigButtInv2} and \ref{FigButtInv4}.
Since this map is regular at the midpoint $M$, which in the upper half
plane coordinates is located at $z=i$ the associated surface is
regular at this point. This is also seen at the pictures
Fig. \ref{FigButtInv2} and \ref{FigButtInv4} but is manifestly not
true for the surfaces associated to butterfly states at
Fig. \ref{FigButt2} and \ref{FigButt4}. The midpoint $M$ is the only point
where the wedge corresponding to $\bra{3}$ has a singularity. Gluing
the wedge with the `wrong sign' butterfly surface associated to the
map (\ref{wrongbuttmap}) results in a surface with simple singularities
which are just cuts and conical singularities which can be smoothed out.
This proves that the state (\ref{ButtInvWedge}) is finite and well
defined state and concludes thus the proof.

On the other hand, changing the sign in the exponent of the butterfly
states would violate the last argument. We would have to glue two
surfaces, one which has a cut originating at the midpoint $M$ and
other which has a conical singularity there. This is a source of
divergences which cancel the vanishing factors in
(\ref{reparU3}). For instance the product like $e^{\frac{1}{2} L_{-2}} \ket{0}
* e^{\frac{1}{2} L_{-2}} \ket{0}$ is finite and well defined but does
not lead to $e^{\frac{1}{2} L_{-2}} \ket{0}$.

\subsection{Some explicit calculations}
\label{DetailedButt}

Using the conservation laws (\ref{ButConsLaw}) and the methods of
\cite{Schnabl3} we may find
\be\label{Ubutt}
U_r e^{\frac12 \tan^2\gamma L_{-2}} \ket{0} =
\l(\frac{2}{r}\r)^{-\frac{c}{16}}
\l(\frac{\cos\frac{2\gamma}{r}}{\cos\gamma}\r)^{-\frac{3c}{16}}
\l(\frac{\sin\frac{2\gamma}{r}}{\sin\gamma}\r)^{\frac{c}{16}}
e^{\frac12 \tan^2 \frac{2\gamma}{r} L_{-2}} \ket{0}.
\ee
We have written the formula for general central charge $c$, now we
shall restrict to the case $c=0$ however. In general, if we know
explicitly the surface state
$U_f U_g^\dagger\ket{0} =  U_{\phi[f,g]}^\dagger\ket{0}$ we may
calculate $U_g U_f^\dagger\ket{0} =  U_{\phi[g,f]}^\dagger\ket{0}$
using
\be\label{phigf}
\phi[g,f] = I \circ \phi[f,g] \circ I \circ f \circ I \circ
g^{-1} \circ I.
\ee
This can be easily established by calculating $\bra{0} \Psi(z) U_f
U_g^\dagger\ket{0}$ for an arbitrary local field $\Psi(z)$. There are
two different ways of bringing it to the form $\aver{f_{1,2}\circ
\Psi(z)}$ whose equivalence requires (\ref{phigf}).

Note that with our definition the functions $\phi[f,g]$ and
$\phi[g,f]$ are defined modulo $SL(2,\cc)$ transformation acting from
the left, which is immaterial for the definition of
the surface states. There is one preferred choice however and this is
the reason why the first factor on the right hand side of (\ref{phigf})
is $I$. With this choice we have
\be\label{UUdag}
U_f U_g^\dagger = U_{\phi[f,g]}^\dagger U_{\phi[g,f]}
\ee
as the operator statement and for the corresponding maps
\be
f \circ I \circ g^{-1} \circ I = I \circ \phi[f,g]^{-1} \circ I \circ
\phi[g,f].
\ee
The equation (\ref{UUdag}) has been used in \cite{SchwarzSen} for the
algebraic proof of the gluing theorem. The maps $\phi[f,g]$ and
$\phi[g,f]$ were constructed as series in an auxiliary parameter
$\alpha$ which controls the contribution of higher levels and which
has to be sent to $1$ at the end of the calculation.
Taking
\be
f(z) = \tan\l(\frac{2}{r}\arctan z\r), \qquad g(z) =
\frac{z}{\sqrt{1-\tan^2\gamma\, z^2}}
\ee
and using (\ref{Ubutt}), (\ref{phigf}) and (\ref{etoL2}) we find easily
\bea
\label{phifgappl} \phi[f,g](z) &=& \frac{z}{\sqrt{1-\tan^2
\l(\frac{2\gamma}{r}\r) \, z^2}}
\\
\label{phigfappl} \phi[g,f](z) &=&
\sqrt{\tan^2\l(\frac{2}{r}\arctan\sqrt{z^2+\tan^2\gamma}\r)  -
\tan^2\frac{2\gamma}{r}}.
\eea

Specializing to $\gamma=\frac{\pi}{4}$ and $r=3$ we find by expanding
(\ref{phigfappl}) in series and solving iteratively for the generating
vector field as in \cite{RZ}
\be
e^{\frac{1}{2}\, L_{2}} U_3^\dagger \ket{0} =
e^{\frac{-3+\sqrt{3}}{12} L_{-2} + \frac{1}{144} L_{-4} +
\frac{-27+7\sqrt{3}}{1552} L_{-6} + \frac{3-\sqrt{3}}{7776}
L_{-8}+\cdots} \ket{0}.
\ee
This supports our claim that it is a well defined state with finite
coefficients and therefore the butterfly state
$e^{-\frac{1}{2}L_{-2}}\ket{0}$
is a true projector.

Let us now look at what happens to the wrong sign butterfly states
under the star product. The function (\ref{phigfappl}) clearly ceases
to be holomorphic at $z=0$ in the limit $\gamma \to i\infty$ for
$r=3$ and the coefficients of the corresponding surface state diverge.
Multiplying the state by the middle factor in (\ref{etoK2k}) which is now
\be
\l(1-(\tan\gamma)^4\r)^{\frac{L_0}{2}} \approx \l(\frac{\sqrt{2}}{\cosh|\gamma|}\r)^{L_0}
\ee
has the effect of changing $z \to \frac{\sqrt{2}z}{\cosh|\gamma|}$ in
(\ref{phigfappl}), which in our limit keeps the domain of
holomorphicity around zero from shrinking to zero size.
Up to an irrelevant overall factor the
function (\ref{phigfappl}) with this substitution has a well defined
limit
\be
\sqrt{\l(1+2z^2\r)^{\frac23} -1 }.
\ee
Finally multiplying the state with $e^{\frac{1}{2} L_{-2}}$ gives a
surface state associated to the map
\be
\sqrt{\l(\frac{1+z^2}{1-z^2}\r)^{\frac23} -1}.
\ee
Using the recursive relations we determine again the generating vector
field and the coefficients in front of the Virasoro generators. The
result is
\be\label{wrongbuttprod}
e^{\frac{1}{2} L_{-2}}\ket{0}*e^{\frac{1}{2} L_{-2}}\ket{0} =
e^{\frac{1}{3} L_{-2} + \frac{5}{54} L_{-4} -
\frac{5}{162} L_{-6} + \frac{59}{2916}
L_{-8}+\cdots} \ket{0}.
\ee
Results from independent numerical check at level 16 are
summarized in Appendix \ref{LevelExpProd}. Note also that this state 
is annihilated by $K_2$ just as the false butterfly $e^{\frac{1}{2}
L_{-2}}\ket{0}$ itself. If we keep multiplying these states we get a
whole commutative subalgebra annihilated by $K_2$, which resembles in many respects the
subalgebra of wedge states, which are characterized by the fact, that
they are annihilated by $K_1$. 

The calculation we have just performed can be easily generalized.
From the formulas (\ref{etoK2k}) and  (\ref{phigfappl}) we can 
calculate the star product of regularized butterflies. The star
product
\be
e^{-\frac12 \tan^2\gamma L_{-2}} \ket{0} * e^{-\frac12 \tan^2\gamma L_{-2}} \ket{0}
= U_{f_{3,\gamma}}^\dagger \ket{0}
\ee
is a surface state associated to the map
\be\label{f3gamma}
f_{3,\gamma}(z)=\sqrt{\tan^2\l(\frac{2}{3}\arctan\sqrt{\frac{z^2+\tan^2\gamma}{1+z^2
\tan^2\gamma}}\r)  - \tan^2\frac{2\gamma}{3}}
\ee
and therefore it looks as
\be
U_{f_{3,\gamma}}^\dagger \ket{0} = e^{v_2 L_{-2} + v_4 L_{-4} + \cdots}
\ket{0},
\ee
where the first two coefficients are 
\bea
v_2 &=& \frac{-3-4 \cos\frac{4\gamma}{3} +
2\cos\frac{8\gamma}{3}}{3 \l(1+2 \cos\frac{4\gamma}{3} \r)^2},
\nonumber\\
v_4 &=&  \frac{\l(1-2 \cos\frac{4\gamma}{3} \r)^2 \l(47+ 60
\cos\frac{4\gamma}{3} + 10 \cos\frac{8\gamma}{3}\r)}{54 \l(1+2
\cos\frac{4\gamma}{3} \r)^4}.
\eea
The multiple star products 
\be
\underbrace{e^{-\frac12 \tan^2\gamma L_{-2}} \ket{0} * e^{-\frac12 \tan^2\gamma L_{-2}} \ket{0}
* \cdots * e^{-\frac12 \tan^2\gamma L_{-2}} \ket{0}}_{(r-1)\,\,
\mbox{\small times}}   
= U_{f_{r,\gamma}}^\dagger \ket{0}
\ee
correspond to maps which look just as (\ref{f3gamma}) with the two factors
of $3$ replaced by $r$. From the formula then follows that the limit
of the repeated multiplication of $e^{\frac{1}{2}
L_{-2}}\ket{0}$ with itself is just the true buttefly state $e^{-\frac{1}{2}
L_{-2}}\ket{0}$.

\subsection{Near butterfly limit}

Let us consider now in detail the limit $\gamma \sim \frac{\pi}{4}$,
in which the state $e^{-\frac12 \tan^2\gamma L_{-2}} \ket{0}$
approaches the butterfly projector.
The product of two such states is a surface state associated to the map
\be
f(z) = \l(\frac{4}{3}\r)^{\frac{5}{4}}
\sqrt{\frac{\pi}{4}-\gamma}\l[\frac{z}{\sqrt{1+z^2}}+
\frac{2}{\sqrt{3}} \l(\gamma- \frac{\pi}{4}\r)
\frac{z}{\l(1+z^2\r)^{\frac{3}{2}}} + O\l(\gamma-
\frac{\pi}{4}\r)^2\r].
\ee
Due to the fact that the second term in the bracket is just a
change of the first under infinitesimal scaling proportional to
\be
z\frac{d}{dz} \l(\frac{z}{\sqrt{1+z^2}} \r) =
\frac{z}{\l(1+z^2\r)^{\frac{3}{2}}}
\ee
we find easily
\bea
e^{s L_{-2}} \ket{0} * e^{ s L_{-2}} \ket{0} &=&
\l[ 1- \l(1-\frac{1}{\sqrt{3}}\r)\l(s+\frac{1}{2}\r) L_{-2} \r] e^{ s
L_{-2}} \ket{0} + O\l(s+\frac{1}{2}\r)^2
\nonumber\\
&=&
\l[ 1+ \l(1-\frac{1}{\sqrt{3}}\r)\l(s+\frac{1}{2}\r) L_{0} \r] e^{ s
L_{-2}} \ket{0} + O\l(s+\frac{1}{2}\r)^2.
\eea
We thus see that $\ket{\psi}= e^{s L_{-2}} \ket{0}$ solves the
equation of motion of ghost number zero \sft
\be
\ket{\psi}*\ket{\psi} = \l(1+a^{-1} L_0\r) \ket{\psi}
\ee
in the limit $a \to \infty$, provided we identify
\be\label{as}
a^{-1} = \l(1-\frac{1}{\sqrt{3}}\r)\l(s+\frac{1}{2}\r).
\ee
This can be directly extended to the case of standard regularized
VSFT, where up to terms of order $O\l(s+\frac{1}{2}\r)^2$ the solution is
\be
\ket{\Psi_0} = \nnn\, e^{s(L_{-2}^X + {L'}_{-2}^{gh})} c_1
\ket{0}
\ee
with the relation between $a$ and $s$ being the same as in (\ref{as}).
The normalization factor $\nnn$ is for $s\ne -\frac12$ finite and calculable.
We have thus fully proven the conjecture of \cite{GRSZ} for which we gave a lot
of evidence in section \ref{ConstrRegVSFT}.

\subsection{Example of divergent star products}
\label{Breakdown}

Now we would like to prove our claim from section \ref{ConstrVSFT}
that the star product
$\ket{B} * L_{-2}^X \ket{B}$ is ill defined. To demonstrate it, we
shall calculate the even level part of the regulated expression
\be
L_{-2}^X e^{\alpha K_2}\ket{0} * e^{\alpha K_2}\ket{0}
\ee
and show that it is divergent in the limit $\alpha \to \infty$.
From (\ref{ButConsLaw}) follows
\be
L_{-2}^X e^{\alpha K_2} \ket{0} =
-\cosh^2 2\alpha \l( K_2^X + \frac{c}{4} \tanh 2\alpha\r)
e^{\alpha K_{2}}\ket{0}.
\ee
Using the $K_2^X$ conservation law and (\ref{transfstar}) we get
\bea
&& L_{-2}^X e^{\alpha K_{2}}\ket{0} * e^{\alpha K_{-2}}\ket{0}
+e^{\alpha K_{-2}}\ket{0} *L_{-2}^X e^{\alpha K_{-2}}\ket{0} =
\nonumber\\
&& \qquad = -\cosh^2 2\alpha \l[ K_{2}^X   + \frac{c}{2} \tanh
2\alpha - 3k_2 \r] e^{\alpha K_{2}} U_3^\dagger\ket{0}.
\eea
Using the relation (\ref{reparU3}) we arrive to
\bea
 \lim_{\alpha\to\infty} \frac{1}{\cosh^2 2\alpha} \l[ L_{-2}^X\,
e^{-\frac{1}{2}\tanh 2\alpha L_{-2}}\ket{0} \r. \!\!\! &*& \!\!\!
e^{-\frac{1}{2}\tanh 2\alpha L_{-2}}\ket{0} +
\\\nonumber
 e^{-\frac{1}{2}\tanh 2\alpha L_{-2}}\ket{0} \!\!\! &*& \!\!\! \l.
 L_{-2}^X\,
e^{-\frac{1}{2}\tanh 2\alpha L_{-2}}\ket{0} \r] =
 c \l(-\frac{1}{4} - \frac{65}{9} \r) e^{-\frac{1}{2}
L_{-2}}\ket{0}
\eea
demonstrating that the star product in the square bracket is
indeed divergent in the limit $\alpha \to \infty$.

\subsection{Family of butterfly states}
\label{Family}

To show that also the whole family of butterfly states
\be
\ket{B,r} = U_r^\dagger e^{-\frac{1}{2} L_{-2}}\ket{0}
\ee
are projectors is rather easy. Important observation is that $A$ given by
(\ref{defA}) annihilates the simplest butterfly
\be
A e^{-\frac{1}{2} L_{-2}}\ket{0} = 0,
\ee
thanks to the identity
\be
\sum_{k=1}^\infty \frac{1}{(2k-1)(2k+1)}  = \frac{1}{2}.
\ee
Therefore, using the formula
\be
e^{-\alpha (A-A^\dagger)} = U_{1+e^\alpha}^\dagger
U_{1+e^{-\alpha}}
\ee
from \cite{Schnabl3} we get
\be
\ket{B,r} =  e^{-\log(r-1) (A-A^\dagger)} e^{-\frac{1}{2}
L_{-2}}\ket{0}.
\ee
The exponent $D\equiv A-A^\dagger$ is an exterior derivative in the
algebra (it is a linear combination of $K_n$'s) and therefore one may
use (\ref{transfstar}) to show that all members of the butterfly
family are projectors and all arise as reparametrizations of the vacuum.
Special cases are the $r \to \infty$ limit, which gives the sliver,
and the $r \to 1$ limit leading to an interesting state which resembles
the identity of \cite{EFHM}
\be
\ket{B,1} = \l(\prod_{n=2}^\infty e^{-\frac{2}{2^n} L_{-2^n}}\r)
e^{-L_{-2}}\ket{0}.
\ee
It differs from the identity only in the sign of $L_{-2}$
in the exponent.

\section{On the uniqueness of the projector}
\label{Uniqueness}

In section \ref{Constraints} we have shown that the simplest butterfly
projector
\be\label{simplbutt}
\ket{\Psi_0} = \nnn\, e^{-\frac12 (L_{-2}^X + {L'}_{-2}^{gh})} c_1
\ket{0}
\ee
does obey all the quadratic identities, which should be satisfied
by all classical solutions of the regularized \sf theory in a
Siegel gauge. An immediate question that arises is, to what extent
is this solution unique. We are going to prove, that in the space
of surface states the identities are so constraining that the
solution (\ref{simplbutt}) is indeed a unique solution.

It turns out, that we do not need to study the whole set of
identities, we will see that already in the matter sector alone
the constraint
\be\label{qidsimpl}
\bra{\Psi_0} L_{2k} \ket{\Psi_0} \sim (-1)^k
\ee
is strong enough to disqualify all other surface states
as possible solutions. The symbol $\sim$ means up to a possibly
divergent normalization constant independent of $k$.

Before we start the actual proof, we can easily demonstrate that the
generalized butterfly projectors certainly do not obey the
identities. Indeed, for $n>2$ even
\be
\bra{0} e^{s L_{n}}  L_{m}\, e^{s L_{-n}} \ket{0} = 0
\qquad  \forall m \in \zz \backslash \{\ldots,-n,0,n,2n,3n,\ldots\}
\ee
but
\be
\bra{0} e^{s L_{n}}  L_{n}\, e^{s L_{-n}} \ket{0} =
\frac{c}{12} (n^3-n)\, s\, \l(1-(ns)^2\r)^{-\frac{c}{12}
\frac{n^2-1}{n} -1} \ne 0.
\ee
The reason is that if $m$ is not an integer multiple of $n$, after
normal ordering of the operators all the terms will have their
$L_0$ weight nonzero, equal to $m$ modulo $n$ and therefore the
expectation value will vanish.

\subsection{Testing the identities for surface states}
\label{TestingIds}

Let us now describe a general method of testing the identity
(\ref{qidsimpl}) applicable to any surface state
\be
\ket{\Psi_0} = U_f^\dagger \ket{0}
\ee
associated to a given conformal map $f(z)$. We are going to calculate
the matrix elements
\be\label{SSEVofT}
\bra{0} U_f L_{-m} U_f^\dagger \ket{0} = \oint \frac{dz}{2\pi i} \frac{1}{z^{m-1}}
\bra{0} U_f T(z) U_f^\dagger \ket{0}.
\ee
From the geometric discussion in section \ref{Butterflies} the state $\bra{0} U_f$,
which formally exists in the Verma module for any $f(z)$
holomorphic near origin, is meaningful in the CFT sense only if
$f(z)$ maps one to one the upper unit half disk into the upper
half plane. The projectors which seem to be characterized by the
fact that $f(i) \to \infty$ are thus limiting cases. The identity
and the false butterflies with $f(\pm 1) \to \infty$ are also
limits of regular surface states. We shall therefore restrict to
the regular surface states only and later discuss these
interesting limits.

We also have to require that
\be
\bra{0} U_f T(z) = \bra{0} f\circ T(z)\, U_f
\ee
is a well defined state in the level expansion. This is not
automatic, since the left hand side is a series in positive and
negative powers of $z$. Negatively moded Virasoro generators when
commuted to the left give rise to infinite series' in mostly
positive powers of $z$, which then should sum to coefficients
\bdm
\frac{(f'(z))^2}{f(z)^{m+2}} \qquad m \in \nn.
\edm
The convergence of these Verma module coefficients follows from
our assumptions about the map $f(z)$ provided $|z|<R$, where $R>1$
is the radius inside which the map is holomorphic and one to one
into the upper half plane. Since we need also $ T(z)
U_f^\dagger\ket{0}$ being well defined, this implies $|z| > 1/R$.
In order for $\bra{0}U_f T(z) U_f^\dagger\ket{0}$ to be absolutely
convergent we need thus
\be
|z| \in (R^{-1},R)  \qquad {\rm and}\quad R>1.
\ee
From what has been said now follows that the right location of the
contour in (\ref{SSEVofT}) is inside the annulus of radii $1/R$ and $R$.
To calculate the integrand we need to use the transformation
properties of the energy momentum tensor $T(z)$
\be
f\circ T(z) = \l(f'(z)\r)^2 T(f(z)) + \frac{c}{12} S[f,z],
\ee
where
\be
S[f,z] = \frac{\partial^3 f \partial f - \frac{3}{2} \l(\partial^2
f\r)^2}{\l(\partial f\r)^2}
\ee
is the Schwarzian derivative.
We can then calculate
\bea\label{UTU}
&& \bra{0}U_f T(z) U_f^\dagger\ket{0} = \bra{0} f\circ T(z) \, U_f
U_f^\dagger\ket{0} = e^{c\kappa} \bra{0} f\circ T(z) \,
U_{\phi[f,f]}^\dagger\ket{0} =
\nonumber\\
&& \quad= e^{c\kappa} \bra{0} I \circ \phi \circ I \circ f\circ T(z)
\ket{0} = \frac{c}{12} S[ I \circ \phi \circ I \circ f,z] \, e^{c\kappa},
\eea
where the function $\phi\equiv\phi[f,f]$ was introduced and
discussed in section \ref{DetailedButt}. The central charge $c$
appearing in the normalization constant $e^{c\kappa} = \aver{0|U_f
U_f^\dagger|0}$ might in principle be different from that of the
energy momentum tensor we insert, but that is just a trivial
extension of our formulas.

Before we move on to the proof, let us have a look on two examples
of the butterfly and sliver projectors.

\subsubsection{Butterfly example}
\label{ButterEx}

For regularized butterfly state associated to the function
\be
f_s(z) = \frac{z}{\sqrt{1-2sz^2}}
\ee
the function $\phi = \phi[f,f]$ is given by
\be
\phi(z) = \frac{z}{\sqrt{(1-4s^2) -2s z^2}}.
\ee
and
\be
I \circ \phi \circ I \circ f(z) = \sqrt{\frac{z^2 -2s}{1 -2s z^2}}.
\ee
The Schwarzian derivative is then
\be
S[I \circ\phi\circ I \circ f, z] =
-\frac{3}{2z^2} \l( 1+ \frac{1-4s^2}{\l(z^2-2s\r)^2}\r)
\l( 1- \frac{1-4s^2}{\l(1-2s z^2 \r)^2}\r).
\ee
We thus see that the contour integral in (\ref{SSEVofT}) receives
contribution from poles at $0, \pm \sqrt{2s}$.
Up to a normalization constant $\aver{0|U_f U_f^\dagger|0}$ it is equal to
\be
\frac{c}{12} \oint \frac{dz}{2\pi i} \frac{1}{z^{2k-1}}
 S[I \circ\phi\circ I \circ f, z] =
 c\, (2s)^k \l(\frac{s^2}{1-4s^2} +\frac{k+1}{8} \r),
\ee
for $m=2k$ even. For $m$ odd the integral vanishes. Although the
above calculation is rather straightforward, these results can be
obtained in a much simpler way by working directly with the
Virasoro generators and conservation laws as we discuss in the
appendices \ref{AppTransf} and \ref{RegGRSZ}. Notice also that in
the limit $s \to -\frac12$ the integral behaves like $(-1)^k$ and
we readily confirm the identities (\ref{qidsimpl}).

\subsubsection{Sliver example}

The situation for the sliver is more complicated. Even if we
regularize the sliver by considering wedge state associated to the
map
\be\label{wedgemap}
f_r(z) = \tan\l(\frac{2}{r} \arctan z\r)
\ee
for large but finite $r$, this function still has branch cuts
starting at $\pm i$ and going all the way to infinity. Therefore
the expression $\bra{0}U_f T(z) U_f^\dagger\ket{0}$ does not
absolutely converge for any values of $z$. In the limit $r \to
\infty$ we can however circumvent this problem quite unambiguously
as we shall see.

The Schwarzian derivative for the wedge states reads
\be
S[f_r,z] =  \frac{4-r^2}{r^2}\, \frac{2}{(1+z^2)^2}
\ee
and the function $\phi_r = \phi[f_r,f_r]$ is given by
\be
\phi_r(z) = f_{4-\frac{4}{r}}(z),
\ee
as was found in \cite{Schnabl3}.
Let us start by calculating
\be
\bra{0} U_f L_{-m} = \frac{c}{12} \oint \frac{dz}{2\pi i} \frac{1}{z^{m-1}}
S[f,z] \bra{0} U_f
+ \oint \frac{dz}{2\pi i} \frac{1}{z^{m-1}} (f'(z))^2 \bra{0} T(f(z))
U_f.
\ee
The contour encircles the origin and remains inside the unit
circle. Then we contract the expression with $U_f^\dagger \ket{0}$.
Up to the overall factor $\bra{0} U_f U_f^\dagger \ket{0}$ the first term gives
\be\label{slivid1term}
\frac{c}{12} \oint \frac{dz}{2\pi i} \frac{1}{z^{m-1}}
S[f,z],
\ee
which is fine as far as concerns the integral. The second term is
then
\be
\frac{c}{12} \oint \frac{dz}{2\pi i} \frac{1}{z^{m-1}}
\frac{(f'(z))^2}{f(z)^4}
S[\phi, I\circ f(z)].
\ee
Note that the form of both terms can be understood as a consequence of
the identity for the Schwarzian derivative
\be
S[I \circ\phi\circ I \circ f, z] = S[f, z] + \frac{(f'(z))^2}{f(z)^4}
S[\phi, I\circ f(z)],
\ee
but we have chosen the stepwise derivation.
The location of the contour in the second term is ambiguous.
In fact it cannot be
deformed in any way to ensure the absolute convergence. We may take it
along the unit circle, but then have to assume some additional
convenient regularization for all $r$. What saves us in the case of
sliver is that
\be
 \frac{(f'(z))^2}{f(z)^4}
S[\phi, I\circ f(z)] =
- 2  \frac{(r-2)(3r-2)}{r^2 (r-1)^2} \l( \frac{1}{1+z^2} \r)^2
\ee
uniformly vanishes for $r \to \infty$. Whatever $r$ independent
regularization we decide for, the second term in the limit will not
contribute. Therefore a unique contribution
comes from the first term (\ref{slivid1term}) and we find
for the sliver $\ket{\Xi}=\lim_{r\to\infty} U_r^\dagger \ket{0}$
\be\label{sliverid}
\bra{\Xi} L_{2k} \ket{\Xi} =
\frac{c}{6}\cdot k (-1)^k \aver{\Xi|\Xi},
\ee
which is clearly different from (\ref{qidsimpl}). The normalization
factor $\aver{\Xi|\Xi}$, which is divergent has been also calculated in
\cite{Schnabl3}.

\subsection{Proof of uniqueness among surface states}

We shall start by rewriting the set of quadratic identities
(\ref{anids}) or (\ref{anidsregVSFT}) in a compact form. Consider
a variation
\be\label{varcompact}
\delta\Psi = \l( T(z) - \frac{1}{z^4} T(-1/z) \r) \Psi_0
\ee
around a solution $\Psi_0$ of the equations of motion of a given \sf
theory. Here $T(z)$ stands for
any of the matter, ghost, twisted ghost energy momentum tensors or
combinations thereof.
For any of these choices we have
\be
[L_0^{tot},T(z)] = \l(2+ z\frac{d}{dz}\r) T(z).
\ee
The cubic term in the \sft action is
left invariant under the variation (\ref{varcompact}). This can be
understood in two ways: First, this variation is an infinite sum of 
variations (\ref{variations}), in which the anomalies happen to cancel
for a natural choice of regularization. Second, the invariance of the
cubic term can be derived from the overlap equations
for the three vertex \cite{HM}  
\be\label{overlap}
\l[T^{(s)}(z) - \frac{1}{z^4} T^{(s+1)}(-1/z) \r] \ket{V_{3}} = 0.
\ee
They are strictly speaking true only for $|z|=1, z\ne\pm i$ and with the 
prescription that in the first term $z$ approaches the unit circle
from outside, whereas in the second term from interior. 
When contracted with regular states, these stringent conditions can be 
relaxed to some annulus around the unit circle.

As a consequence of the equations of motion, if the cubic term is
invariant under the variation, so has to be also the kinetic term.
From that we find an important relation
\be
\l(2+ z\frac{d}{dz}\r) \bra{\Psi_0} T(z) \ket{\Psi_0} = 0,
\ee
whose unique solution is
\be\label{compid}
\bra{\Psi_0} T(z) \ket{\Psi_0} = \frac{\lambda}{z^2}.
\ee
This equation contains as much information as the original set
of identities (\ref{anids}) or (\ref{anidsregVSFT}).
The constant $\lambda$ depends on the type of the energy momentum
tensor and is also related to the anomaly. Details depend also on
whether we are in vacuum or ordinary \sf theory.

Now we are going to solve the equation (\ref{compid}) in the space of
surface states discussed in section \ref{TestingIds}.
Due to (\ref{UTU}) our problem reduces to the equation
\be\label{Sgeq}
S[I \circ \phi \circ I \circ f,z]  = \frac{\mu}{z^2},
\ee
where we denote
\be
\mu = \frac{12\lambda}{c}\, e^{- c\kappa}.
\ee
Although the equation (\ref{Sgeq}) looks as a third order nonlinear
differential equation we can find easily all its solutions. Denoting
\be
g(z) = I \circ \phi \circ I \circ f(z)
\ee
we find the solution
\bea\label{gsol}
g(z) &=& h\l( z^{\sqrt{1-2\mu}} \r)  \qquad {\rm for} \quad \mu \ne \frac{1}{2},
\\
g(z) &=& h\l( \log z \r)  \qquad {\rm for} \quad \mu = \frac{1}{2},
\eea
where $h(z)$ is an arbitrary $SL(2,\rr)$ map. Note that for this
reason the sign of the square root does not matter. The map $h(z)$
is parameterized by three real parameters, just as many as
the number of parameters necessary to specify the initial
conditions of a third order differential equation.

So we have managed to reduce the problem to the equation
\be\label{phiIfeq}
I \circ \phi \circ I \circ f(z) = h \circ z^n,
\ee
restricting for simplicity to the case $\mu \ne \frac{1}{2}$  and
denoting $n=\sqrt{1-2\mu}$. Now, a bit surprisingly the equation
(\ref{phiIfeq}), for $n$ even at least, does not admit any 
regular solution.  To see that, note
that from (\ref{phigf}) the left hand side of (\ref{phiIfeq}) 
commutes with the inversion
\be
I \circ (I \circ \phi \circ I \circ f) = (I \circ \phi \circ I \circ f) \circ I,
\ee
but the right hand side never commutes, due to the fact that 
\be
I \circ h = h \circ I \circ (-{\rm id})
\ee
can be never satisfied within $SL(2,\rr)$ group.
What we can do however, is to find a sequence of approximations $g_{n,a}(z)$
for $a \to 1$ whose limit is $g(z)= h \circ z^n$ and hope that the 
corresponding sequence of $f_{n,a}$'s whose uniqueness we shall prove,
have always the same limit. This is just a technical assumption of 
smooth dependence of $f$ on $g$.

The most natural sequence of approximations labelled with a
continuous parameter $a$ is
\be\label{gndef}
g_{n,a}(z) = \l(\frac{z^n+(-1)^n a}{1+a z^n}\r)^{\frac{1}{n}}.
\ee
For $n$ even and $ a = 1+\eps$ we have
\be
g_{n,a}(z) = 1 +  \frac{\eps}{n} \cdot \frac{1-z^n}{1+z^n} + O(\eps)^2
\ee
and this has indeed the form of $h \circ z^n$ up to the first order
in $\eps$. We would have obtained a similar result also for $a$ close to $-1$.
The Schwarzian derivative (for all $n$) is
\be
S[g_{n,a},z] = - \frac{n^2-1}{2z^2} \l(1+ \frac{1-a^2}{(z^n+(-1)^n a)^2}\r)
\l(1- \frac{1-a^2}{(1+a z^n)^2}\r)
\ee
and in the limit $a \to \pm 1$ it clearly approaches $-\frac{n^2-1}{2z^2}$.

Now we are going to  determine the surface states, which actually solve our
equation (\ref{phiIfeq}) with $h\circ z^n$ replaced by $g_{n,a}(z)$. 
There are two ways to proceed.
Based on our results from subsection \ref{ButterEx}, we could guess
and check that (\ref{gndef}) requires
\be
f(z) = \frac{z}{\l(1 + a z^n\r)^{\frac{1}{n}}}
\ee
and the limits $a \to \pm 1$ correspond
to the butterfly and false butterfly states. 

What about the uniqueness? It seems that we are getting more solutions
depending on $n$, recall however, that different $n$ correspond to
different $\lambda$ which we take fixed in such a way, that the butterfly 
state $e^{-\frac{1}{2} L_{-2}} \ket{0}$ solves the constraint.
Are there any other solutions? We can prove uniqueness of the solution
of the equation 
\be
I \circ \phi \circ I \circ f(z) = g(z) 
\ee
for any given $g(z)$ by the following reasoning: 
Suppose there is another solution
$\tilde f$ with $\tilde\phi=\phi[\tilde f,\tilde f]$ such that
\bdm
I \circ \phi \circ I \circ f = I \circ \tilde\phi \circ I \circ
\tilde f.
\edm
Since the operator corresponding to
$I \circ \phi \circ I$ contains only negatively moded Virasoro
operators, we find $\bra{0} U_f = \bra{0} U_{\tilde f}$ and thus
$f(z) = \tilde f(z)$ modulo some unimportant global rescaling.

An alternative procedure is to recall from \cite{LPP1} that the
map (\ref{gndef}) is
generated by a vector field $v(z) = z^{n+1} - (-)^n z^{-n+1}$, which obeys
$v(Iz) = \frac{1}{z^2} v(z)$.\footnote{In general any map $g$
satisfying $g\circ I = I\circ g$ is generated by a vector field $v(z)$ 
obeying $v(Iz) = \frac{1}{z^2} v(z)$, as can be seen from the equation 
$v(z) \partial g(z) = v(g(z))$.}
The operator which corresponds to the
map $g_n$ is thus
\be
U_{g_n} = e^{\alpha K_n}.
\ee
The dependence of $\alpha$ on $s$ should be worked out separately.
Again, since the operator corresponding to
$I \circ \phi \circ I$ acts as identity on the vacuum $\bra{0}$
we have
\be
\bra{0} U_f = e^{\frac{c\kappa}{2}} \bra{0} e^{\alpha K_n},
\ee
where $e^{c\kappa} = \bra{0} U_f U_f^\dagger \ket{0}$.
From the formulas in Appendix \ref{AppGroup} we easily find
the relation between the parameters $\alpha$ and $s$
\be
s = \frac{1}{2k}\tanh 2k\alpha.
\ee

As a byproduct of the above arguments we obtain an interesting statement.
For any regular surface states $\bra{0} U_f$ associated to the
maps specified above,
there exists a star algebra derivative $K$, which is a linear
combination of the generators $K_n=L_n - (-1)^n L_{-n}$, such that
\be
e^K =  e^{-\frac{c\kappa}{2}} U_{\phi^{-1}}^\dagger U_f =
e^{\frac{c\kappa}{2}} U_{\phi} U_{f^{-1}}^\dagger
\ee
and therefore
\be\label{byprod}
\bra{0} U_f = e^{\frac{c\kappa}{2}} \bra{0} e^K.
\ee
This follows from the fact, that the function $g=I \circ \phi \circ
I\circ f$ is generated by some $K$ and from the fact that $I \circ \phi \circ
I\circ$ is holomorphic around $\infty$ and hence the corresponding
operator acts trivially on $\bra{0}$.
The equation (\ref{byprod}) has been checked on a number of examples
like wedge states in \cite{Schnabl3} and regularized butterflies in
the present paper.

\section{Conclusions}

We have seen that the anomalous $K_n$ symmetries, which are just 
consequences of the overlap equations for the string vertex,
yield nontrivial information about the solutions of the ordinary or
vacuum \sf theories. 

In order to obtain finite action solutions in VSFT
with pure ghost kinetic operator, we generically need infinite
normalization of the action. Due to the singular nature of projectors 
it could happen that the normalization turns out to be finite. In any
case it shows that some sort of regularization of the string field
action is inevitable.

Solutions of the regularized version of VSFT constructed by adding a
term $a^{-1} c_0 L_0$ to the kinetic operator obey such a constraint,
which among all the surface states admits unique solution which is
just the simplest butterfly projector. The constraint, up to a single
constant, is the same as for the ordinary open \sft (OSFT). In that 
case we know however, that we should look for the solution of the full
equations of motion outside the family of surface states. 

The overlap equations encode all the information about the \sft vertices.
It would be interesting to see if we could derive more identities
based on overlap equations for products of local operators. Such
identities could then uniquely specify all the coefficients 
in the solution \cite{SZ,MT}, except for a single parameter.

In a recent paper \cite{GRSZ2} by Gaiotto et al. the authors found an 
intriguing coincidence for the OSFT solutions. The coefficients in the
matter sector agree within few percent with coefficients of a solution
of a ghost number zero \sf theory, whose
equation of motion in the Siegel gauge reads 
\be
(L_0-1)\ket{\Psi}= - \ket{\Psi}*\ket{\Psi}.
\ee
Even though the solution is given purely in terms of total Virasoro
operators acting on the vacuum, we can still find our quadratic 
identities based on $K_n$'s in the matter sector. They happen to be
the same as the matter part of such identities in the OSFT. Whether
one can disentangle the matter and ghost sectors and add more
identities to fully prove the observed pattern remains to be seen.
   
Another issue discussed in the paper are the surface states and projectors. 
We have shown that $K_n$ star algebra derivatives can be useful to
calculate some nontrivial star products and to find some projectors.
It would be interesting to obtain full classification of all projectors
and to clarify their relation to star algebra derivatives in general.

\section*{Acknowledgements}
I would like to thank L.~Rastelli, A.~Sen, W.~Taylor
and B.~Zwiebach for useful discussions. This work has been
supported in part by DOE contract \#DE-FC02-94ER40818.

\newpage

\appendix

\section{$SL(2,\rr)$ group identities}
\label{AppGroup}

Let us take an $sl(2,\rr)$ algebra with a central extension satisfying
the commutation relations
\bea
\l[T_1,T_{-1}\r] &=& 2 T_0 + a,
\nonumber\\
\l[T_0,T_1\r] &=&  -T_1,
\nonumber\\
\l[T_0,T_{-1}\r] &=&  T_{-1}.
\eea
Then we have
\bea
e^{\alpha(T_1-T_{-1})} &=& e^{-\tanh\alpha\, T_{-1}}\,
 \l(\cosh\alpha\r)^{-( 2 T_0 + a)} \, e^{\tanh\alpha\, T_{1}},
\nonumber\\
e^{\alpha(T_1+T_{-1})} &=& e^{\tan\alpha\, T_{-1}}\,
 \l(\cos\alpha\r)^{-( 2 T_0 + a)} \, e^{\tan\alpha\, T_{1}},
\nonumber\\
e^{\alpha T_1} \, e^{\beta T_{-1}} &=& e^{\frac{\beta}{1-\alpha\beta} T_{-1}}\,
(1-\alpha\beta)^{-( 2 T_0 + a)} \, e^{\frac{\alpha}{1-\alpha\beta} T_{1}},
\nonumber\\
e^{\alpha T_1} \, e^{\beta T_{0}} \; &=& e^{\beta T_{0}}\, e^{\alpha\, e^\beta T_1}.
\eea
These identities can be presumably established by a variety of means.
We have found them by constructing explicitly an infinite dimensional representation
built by successive actions of $T_{-1}$ on some vacuum state $\ket{0}$ which is
annihilated by $T_0$ and $T_1$ and then using the method of
differential equations described in \cite{Schnabl3}. Much simpler way
to obtain them is by analytic continuation from analogous equations
for $SU(2)$ group.

When applied to subalgebras of the Virasoro algebra through the identification
\be
T_1=\frac1n  L_n, \quad T_{-1}= \frac1n L_{-n}, \quad T_0=\frac1n L_0
\ee
and
\be
a= \frac{c}{12} \frac{n^2-1}{n},
\ee
we find easily the following important relations
\bea
e^{\alpha K_1} &=& e^{\tan\alpha\, L_{-1}}\,
 \l(\cos\alpha\r)^{-2 L_0 } \, e^{\tan\alpha\, L_{1}},
\\
\label{etoK2}
e^{\alpha K_2} &=& e^{-\frac{1}{2}\tanh 2\alpha\, L_{-2}}\,
 \l(\cosh 2\alpha\r)^{- L_0-\frac{c}{8}} \, e^{\frac{1}{2} \tanh 2\alpha\, L_{2}},
\\
e^{\alpha K_{2k}} &=& e^{-\frac{1}{2k}\tanh 2k\alpha\, L_{-2k}}\,
 \l(\cosh 2k\alpha\r)^{- \frac{1}{k} L_0-\frac{c}{24}\frac{4k^2-1}{k}}
 \, e^{\frac{1}{2k} \tanh 2k\alpha\, L_{2k}},
\eea
where $K_n=L_n -(-1)^n L_{-n}$ as usual.

\newpage
\section{Transformation properties of the butterfly}
\label{AppTransf}

The regularized version of the simplest butterfly state
$e^{s L_{-2}} \ket{0}$ is also a surface state, since as we know
from \cite{LPP2}
\be\label{etoL2}
e^{s L_{2}} = U_{\frac{z}{\sqrt{1-2sz^2}}}.
\ee
By straightforward use of Virasoro commutation relations or
by the general techniques of \cite{RZ} we may
find various conservation laws for this state
\bea\label{ButConsLaw}
(L_0 - 2s L_{-2}) e^{s L_{-2}} \ket{0} &=& 0,
\nonumber\\
(L_{2} - (2s)^{2} L_{-2}) e^{s L_{-2}} \ket{0} &=&
\frac{c s}{2}   e^{s L_{-2}} \ket{0},
\nonumber\\
(L_{2k} - (2s)^{k+1} L_{-2}) e^{s L_{-2}} \ket{0} &=&
\frac{c(k+1)}{8} (2s)^k  e^{s L_{-2}} \ket{0}, \qquad k \ge 1.
\eea
We keep the central charge $c$ arbitrary.
Applying simple method of differential equations (see e.g. \cite{Schnabl3})
one can derive from these conservation laws formulas for the action of
finite conformal transformation
\bea\label{ButFinTransf}
e^{v L_0} e^{s L_{-2}} \ket{0} &=&  e^{s e^{2v} L_{-2}} \ket{0},
\nonumber\\
e^{v L_{2}} e^{s L_{-2}} \ket{0} &=& \l(1- 4vs \r)^{-\frac{c}{8}}
\exp\l(\frac{s}{1-4vs} L_{-2}\r) \ket{0},
\nonumber\\
e^{v L_{2k}} e^{s L_{-2}} \ket{0} &=& \l(1-2kv (2s)^k \r)^{-\frac{c}{16} \frac{k+1}{k}}
\exp\l(\frac{s}{\sqrt[k]{1-2kv(2s)^k}} L_{-2}\r) \ket{0}.
\eea
By the same method, or using the formula (\ref{etoK2}) from Appendix \ref{AppGroup}
as a shortcut, one can derive the action of a finite reparametrization generated by
$K_2=L_2-L_{-2}$
\be\label{etoKetoL}
e^{\alpha K_2} e^{s L_{-2}} \ket{0} = (\cosh 2\alpha - 2s \sinh
2\alpha)^{-\frac{c}{8}} \exp\l(\frac{2s \cosh 2\alpha - \sinh 2\alpha}{2\cosh 2\alpha
- 4s \sinh 2\alpha} L_{-2}\r) \ket{0}.
\ee
Note, that from this formula follows for $c=0$ and $s\in [-\frac{1}{2},\frac{1}{2})$
\be
\lim_{\alpha\to\infty} e^{\alpha K_2} e^{s L_{-2}} \ket{0} =
e^{-\frac 12 L_{-2}} \ket{0}.
\ee
This is quite analogous to (\ref{reparU3}), in that infinite
reparametrization of some set of states yields the butterfly projector.
It can be proven along the same lines since the surfaces corresponding
to $\bra{0}  e^{\frac{1}{2} L_{2}}$ and $\bra{0}  e^{s L_{2}}$
for $s\in [-\frac{1}{2},\frac{1}{2})$ can be glued together without
gluing singularity to singularity.

\section{Some formulas for the regulated GRSZ solution}
\label{RegGRSZ}

Here we collect some useful formulas for the regularized butterfly
states in a twisted conformal field theory
\be
\ket{\Psi^{s, \tilde s}} =  e^{ s L_{-2}^X + \tilde s
L_{-2}'{}^{gh}} c_1 \ket{0}.
\ee
Using (\ref{ButConsLaw}) and  (\ref{ButFinTransf}), one can easily
establish for this state (omitting the superscripts)
\bea
\bra{\Psi} c_0 \ket{\Psi} &=& (1-4s^2)^{-\frac{26}{8}} (1-4{\tilde s}^2)^{\frac{2}{8}},
\nonumber\\
\bra{\Psi} c_0 L_0^X \ket{\Psi} &=& \frac{26 s^2}{1-4s^2} \bra{\Psi} c_0 \ket{\Psi},
\nonumber\\
\bra{\Psi} c_0 {L'}_0^{gh} \ket{\Psi} &=& -\frac{2 {\tilde
s}^2}{1-4{\tilde s}^2} \bra{\Psi} c_0 \ket{\Psi},
\nonumber\\
\bra{\Psi} c_0 L_0^{tot} \ket{\Psi} &=& \l(\frac{26 s^2}{1-4s^2}- \frac{2 {\tilde
s}^2}{1-4{\tilde s}^2} -1 \r)
 \bra{\Psi} c_0 \ket{\Psi},
\nonumber\\
\bra{\Psi} c_0 L_{2k}^X \ket{\Psi} &=& 26 (2s)^k \l(\frac{s^2}{1-4s^2}
+\frac{k+1}{8} \r) \bra{\Psi} c_0 \ket{\Psi},
\nonumber\\
\bra{\Psi} c_0 {L'}_{2k}^{gh} \ket{\Psi} &=& -2 (2\tilde s)^k \l(\frac{{\tilde
s}^2}{1-4{\tilde s}^2}
+\frac{k+1}{8} \r) \bra{\Psi} c_0 \ket{\Psi}.
\eea

\section{Multiplication of butterflies in level expansion}
\label{LevelExpProd}

In this appendix we present some numerical data showing how well the
butterfly projectors behave in level expansion.

We have calculated first three coefficients of the product
$ e^{-\frac{1}{2} L_{-2}} \ket{0} * e^{-\frac{1}{2} L_{-2}} \ket{0}$
up to level 16. Calculation up to level $L$ means, that we truncate 
the input state up to that level and then perform the star
multiplication exactly. The results are
\begin{displaymath}
\begin{array}{|c|c|c|c|}
\hline
{\rm Level} & L_{-2} \ket{0} & L_{-2} L_{-2} \ket{0}  &  L_{-4} \ket{0} \\ \hline
2   & -0.4323 & 0.0937 &   -0.04069 \\ \hline
4   & -0.4603 & 0.1084 &   -0.00351 \\ \hline
6   & -0.4721 & 0.1131 &   -0.00213 \\ \hline
8   & -0.4786 & 0.1157 &   -0.00146 \\ \hline
10  & -0.4826 & 0.1173 &   -0.00108 \\ \hline
12  & -0.4854 & 0.1185 &   -0.00084 \\ \hline
14  & -0.4874 & 0.1193 &   -0.00067 \\ \hline
16  & -0.4889 & 0.1200 &   -0.00055 \\ \hline
\infty  & -0.4955 & 0.1234 &   -0.00019 \\ \hline
{\rm exact}  & -0.5000 & 0.1250 &   0 \\ \hline
\end{array}
\end{displaymath}

For the product
$ e^{\frac{1}{4} L_{-4}} \ket{0} * e^{\frac{1}{4} L_{-4}} \ket{0}$
we have calculated the first three coefficients up to level 24:
\begin{displaymath}
\begin{array}{|c|c|c|c|}
\hline
{\rm Level} & L_{-2} \ket{0} & L_{-2} L_{-2} \ket{0}  &  L_{-4} \ket{0} \\ \hline
4   & -0.0709 & -0.00296 &  0.2418 \\ \hline
8   & -0.0494 & -0.00136 &  0.2452 \\ \hline
12  & -0.0398 & -0.00080 &  0.2466 \\ \hline
16  & -0.0342 & -0.00054 &  0.2473 \\ \hline
20  & -0.0304 & -0.00039 &  0.2478 \\ \hline
24  & -0.0276 & -0.00030 &  0.2482 \\ \hline
\infty  & -0.0211 & 0.00025 & 0.2492 \\ \hline
{\rm exact}  & 0 & 0 & 0.2500 \\ \hline
\end{array}
\end{displaymath}
The values for level $L=\infty$ were obtained in both cases by a simple fit
$a+\frac{b}{L}$. For some coefficients a different power of $L$ seems more
accurate, but due to lack of data we refrain from more careful
analysis. Anyway, the numerical results are compatible with
butterflies being projectors.

Finally as a check on our methods and formulas we present numerical
calculation of the star product (\ref{wrongbuttprod})
\bea
e^{\frac{1}{2} L_{-2}}\ket{0}*e^{\frac{1}{2} L_{-2}}\ket{0} &=&
e^{\frac{1}{3} L_{-2} + \frac{5}{54} L_{-4} -
\frac{5}{162} L_{-6} + \frac{59}{2916}
L_{-8}+\cdots} \ket{0}
\nonumber\\
&=& \l(1 + 0.3333 L_{-2} + 0.0926 L_{-4}  - 0.0309
L_{-6} + 0.0555 L_{-2} L_{-2} + \r.
\nonumber\\
&& \; \l.  + 0.0154 (L_{-2} L_{-4} + L_{-4} L_{-2})
+ 0.00617 L_{-2} L_{-2} L_{-2} +\cdots \r) \ket{0}.
\nonumber
\eea
Numerical calculation at level 16 gives
\bea
e^{\frac{1}{2} L_{-2}}\ket{0}*e^{\frac{1}{2} L_{-2}}\ket{0} &=&
\l(1 + 0.3001 L_{-2} + 0.0928 L_{-4}  - 0.0272
L_{-6} + 0.0452 L_{-2} L_{-2} + \r.
\nonumber\\
&& \;  + 0.0140 (L_{-2} L_{-4} + L_{-4} L_{-2})
+ 0.00388 L_{-2} L_{-2} L_{-2} +
\nonumber\\
&& \; \l.
- 0.00022 L_{-3} L_{-3} + \cdots \r) \ket{0},
\nonumber
\eea
which is at least roughly compatible with the exact answer.

\end{document}